\documentclass[prl,aps,reprint,showpacs,amssymb,superscriptaddress]{revtex4-1}
\usepackage{graphicx}
\newcommand{\kvec}{{\bf k}}
\newcommand{\qvec}{{\bf q}}
\begin{document}
\title{Gutzwiller theory of band magnetism in LaOFeAs}
\author{Tobias~Schickling}
\author{Florian Gebhard}
\affiliation{Department of Physics and Materials Science Center, 
Philipps Universit\"at, 35032 Marburg, Germany}
\author{J\"org~B\"unemann}
\affiliation{Institut f\"ur Physik, BTU Cottbus, P.O.\ Box 101344, 
03013 Cottbus, Germany}
\author{Lilia Boeri}
\author{Ole K.\ Andersen}
\affiliation{Max-Planck--Institute for Solid-State Research,
Heisenbergstr.~1, 70569 Stuttgart, Germany}
\author{Werner Weber} 
\affiliation{Theoretische Physik II,
Technische Universit\"at Dortmund, Otto-Hahn-Str.\ 4,
44227 Dortmund, Germany}

\date{\today}
\begin{abstract}
We use the Gutzwiller variational theory to calculate the ground-state phase
diagram and quasi-particle bands of LaOFeAs. The Fe3$d$--As4$p$ 
Wannier-orbital basis obtained from density-functional theory defines
the band part of our eight-band Hubbard model.
The full atomic interaction between the electrons in the 
iron orbitals is parameterized by the Hubbard interaction~$U$ and an average 
Hund's-rule interaction~$J$. We reproduce the experimentally observed 
small ordered magnetic moment over a large region of $(U,J)$ parameter 
space. The magnetically ordered phase is a stripe 
spin-density wave of quasi-particles.
\end{abstract}

\maketitle

LaOFeAs is the first high-temperature Fe-based
superconductor~\cite{kamihara2008}. Its crystal structure is made
of alternating layers of (LaO)$^{+}$ and (FeAs)$^{-}$.
Below $T_{\rm s}=150\, {\rm K}$, the material undergoes
a structural phase transition from tetragonal to orthorhombic,
followed by the onset of a stripe spin-density wave (SDW) order
in the FeAs layers with wave vector $\qvec=(0,\pi)$.
The values of the ordered moments vary from
$m=0.3\mu_{\rm B}$ to 
$0.8\mu_{\rm B}$~\cite{magmom1-etal,magmom2-etal,magmom3-etal}.
Under pressure or upon doping, the electrons in the FeAs layers
become superconducting 
at  $T_{\rm c} = 56\, {\rm K}$~\cite{FESC:review,FESC:review-2}.
The presence of Fe-pnictogen or chalcogen layers 
and the proximity of magnetism and
superconductivity with fairly high $T_{\rm c}$
are common to almost all the iron-based superconductors known to date, 
and place these compounds in the class of unconventional superconductors
like the cuprates. 

However, the cuprates and pnictides show important differences.
In contrast to the cuprates, the normal phase of the pnictides is 
always metallic; in fact, 
density-functional theory (DFT) calculations 
provide a qualitatively correct description of the 
electronic structure and Fermi surface (FS) of the
paramagnetic (PM) phase~\cite{magmomdft1-lilia,ole-lilia},
and reproduce the correct magnetic structure
of the SDW state~\cite{magmomdft2-lilia-etal}.
However, sizable renormalizations 
of the DFT quasi-particle band masses are observed
in optics, angular-resolved photo-emission spectroscopy (ARPES), 
and de-Haas--van-Alphen measurements~\cite{FESC:review}. Moreover,
the values of the ordered moments calculated by DFT in the SDW phase,
$m_{\rm DFT}\approx 1.8\mu_{\rm B}$ or higher, 
are much larger than the experimentally observed values~\cite{magmomdft1-lilia}.

This discrepancy has lead to an intense ongoing debate on the nature
of magnetism (localized vs.\ itinerant), and on the
mechanism of suppression of the magnetic moment 
(long-range spin fluctuations vs.\ strong local electronic
correlations). Adopting the correlated-electron 
viewpoint~\cite{aichhorn-prb2009-etal,kotliarNPL},
one faces the problem that very few theoretical methods are available
for the description of itinerant magnetism in multi-band systems.

In this work, we employ the Gutzwiller variational theory (GT) 
which describes the ground state and quasi-particle excitations 
of multi-band Fermi liquids
to study the magnetic phase diagram of LaOFeAs;
GT treats the full atomic interactions and
is numerically much less involved than, e.g., 
Dynamical Mean-Field 
Theory (DMFT)~\cite{aichhorn-prb2009-etal,hansmann-etal,kotliarNPL}.
This enables us to resolve the small energy differences 
between the PM and SDW phases in the pnictides. 
We shall find that the striped SDW in LaOFeAs has a small ordered moment
over a large region of $(U,J)$ parameter space.
This small-moment phase can be understood as the band magnetism of 
correlated quasi-particles.

\begin{figure}[tb]
\begin{minipage}{8.65cm}
{\tabcolsep=-6pt\begin{tabular}{@{}ll@{}}
\hspace*{-24pt}\includegraphics[width=4.6cm]{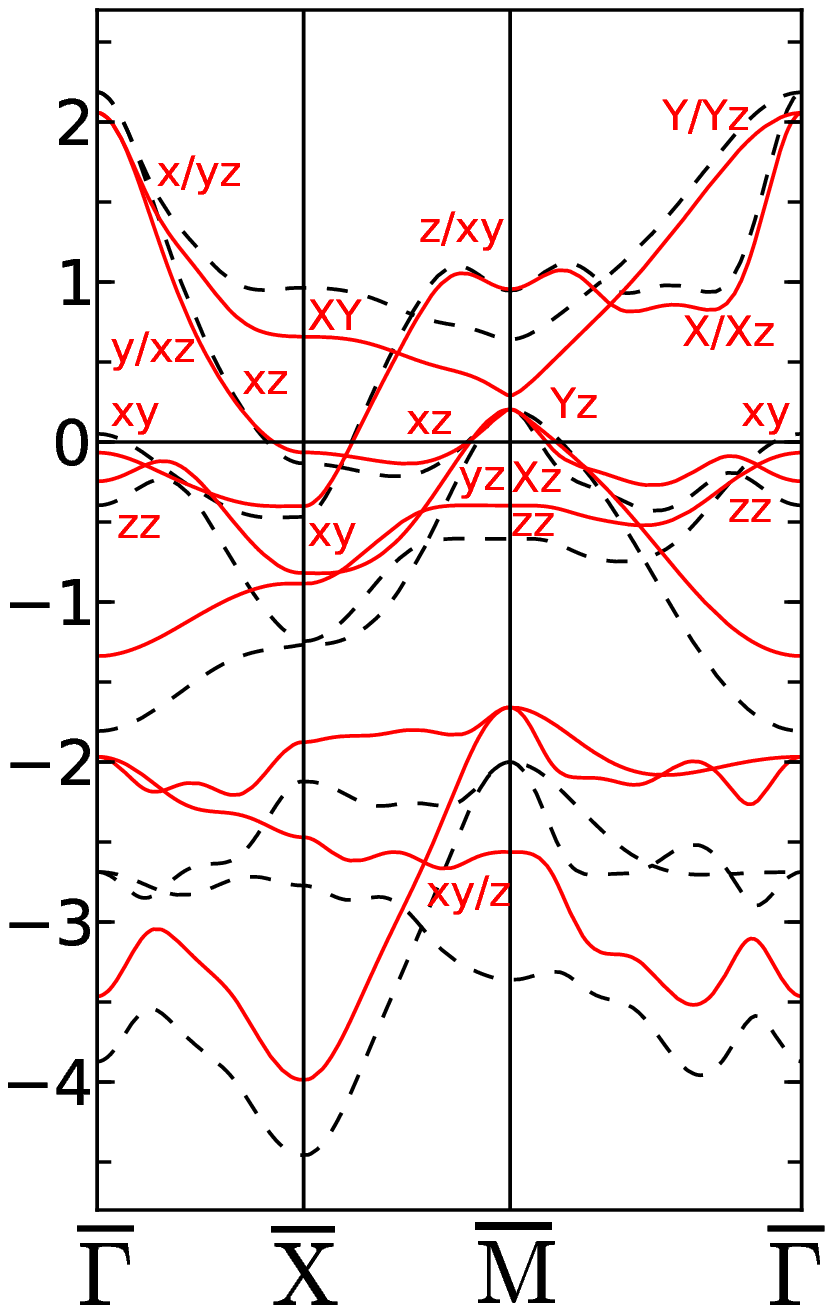}
& 
\includegraphics[width=4.7cm]{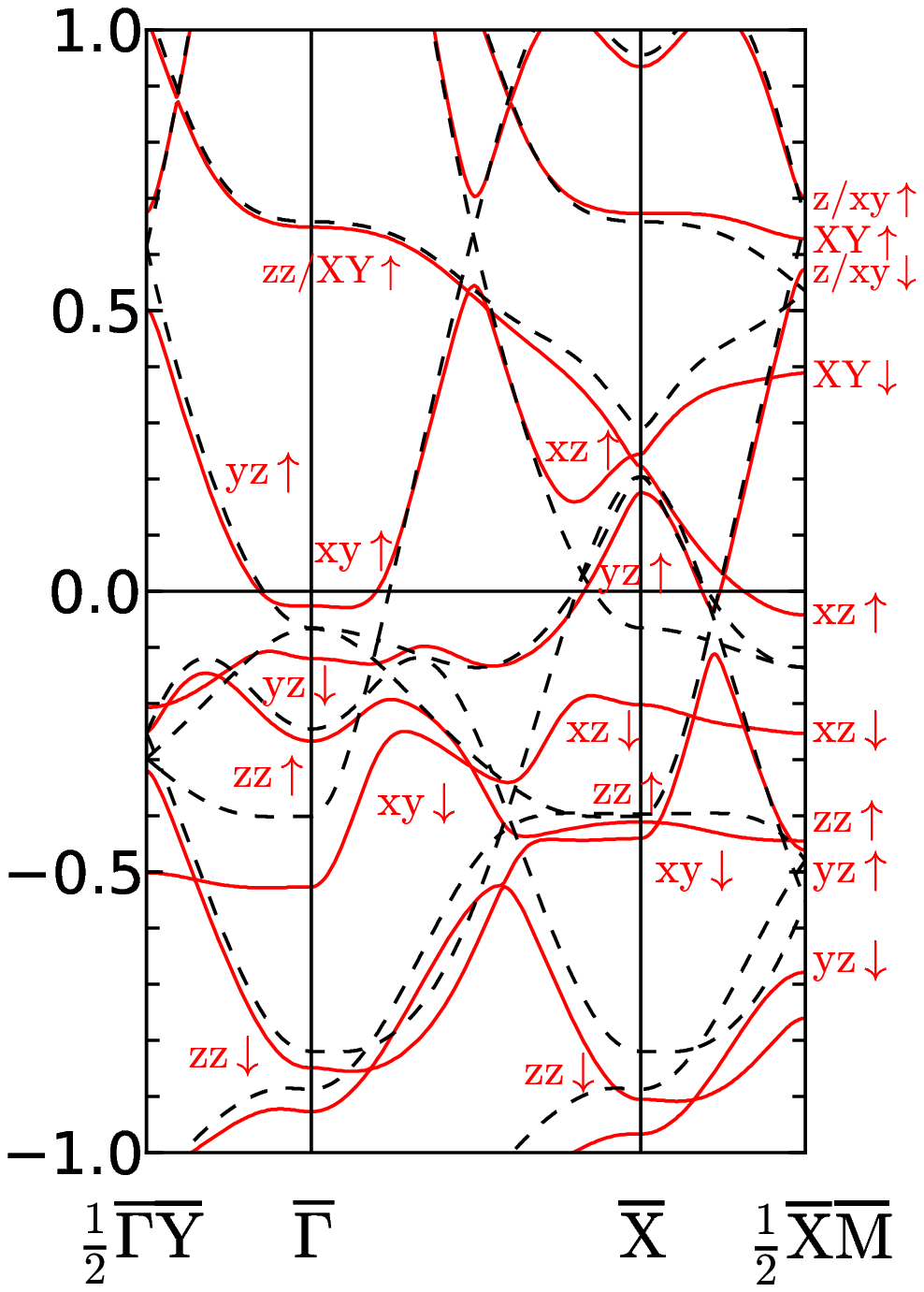}
\end{tabular}}
\end{minipage}
\caption{(Color online) Fe3$d$--As4$p$ bandstructure of LaOFeAs
along the high-symmetry lines of the 2D Brillouin zone
containing one formula unit per cell.
Left: DFT~\protect\cite{ole-lilia} (dashed)
and GT (solid) in the PM phase
for $(U,J)=(8,0.6)\, {\rm eV}$. The bands are lined up at the Fermi level
($E=0$), energies on the ordinate are in eV,
the DFT bandwidth is $W_{\rm DFT}=6.5\, {\rm eV}$.
Right: GT bands for $(U,J)=(8,0.6)\, {\rm eV}$ 
in the PM (dashed) and SDW (solid) phases 
in the Brillouin zone folded along the line 
$\frac{1}{2}\overline{\Gamma}\overline{\rm Y}$-$\frac{1}{2}\overline{\rm X}\overline{\rm M}$.
High symmetry points and the directions $x$, $y$ and $X$, $Y$
are defined in Fig.~\protect\ref{fig:fermisurface}.
Dominant orbital and spin characters are indicated. 
Spin-$\uparrow$ are minority and spin-$\downarrow$ are majority spin states.
For strongly covalent bands,
$z/xy$ means more $z$ and less $xy$ character 
and vice versa~\protect\cite{ole-lilia}.\label{fig:bandstructure}}
\end{figure}

We investigate the two-dimensional eight-band Hubbard model
\begin{equation}
\hat{H}=\hat{H}_0+\hat{H}_{\rm C}=\sum_{i,j;b,b';\sigma}t_{i,j}^{b,b'}                 
\hat{c}_{i,b,\sigma}^{\dagger}\hat{c}_{j,b',\sigma}^{\phantom{\dagger}}        
+\sum_i \hat{H}_{{\rm C},i} \;,
\label{1.1} 
\end{equation}       
where $\hat{H}_0$ describes the on-site energies
and hopping integrals 
of electrons with spin $\sigma=\uparrow,\downarrow$
between Fe3$d$-As4$p$ Wannier orbitals $b,b'$ 
on sites $i,j$ 
(NMTO)~\cite{ole-lilia}. The corresponding DFT bandstructure is shown 
in Fig.~\ref{fig:bandstructure}.
We have neglected the interlayer hoppings and have used the glide plane
to unfold to the Brillouin zone which contains only one formula unit per  
cell~\cite{ole-lilia}.

\begin{figure}[htbp]
\includegraphics[width=6.8cm]{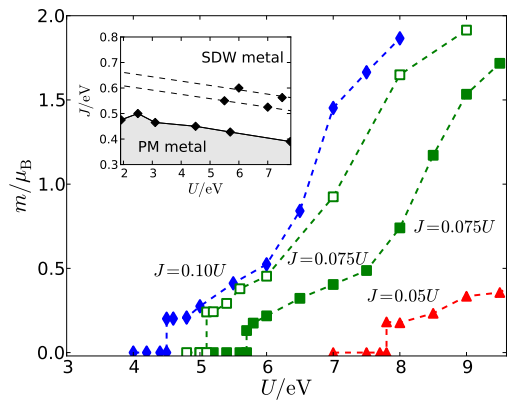}
\caption{(Color online) GT ordered magnetic moment 
as a function of $U$ for various values of $J/U$.
Our characteristic point $(U,J)=(8,0.6)\, {\rm eV}$ is a filled square.
Full symbols: full atomic Hamiltonian; open symbols:
restricted atomic Hamiltonian with density-density interactions only.
Inset: Ground-state phase diagram as a function of $U$ and $J$.
In the SDW metal, we also show lines for constant magnetization, 
$m=0.4\mu_{\rm B}$ and $m=0.5\mu_{\rm B}$, respectively.
\label{fig:magmomGT}}
\end{figure}

\begin{figure}[t]
\includegraphics[height=5.6cm]{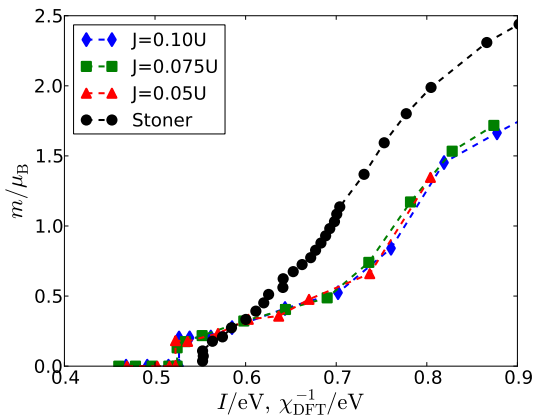}
\caption{(Color online) Ordered moment as a function
of $I=J+0.017U$ for various values of $J/U$ in GT (diamonds),
and the inverse susceptibility in DFT $\chi_{\rm DFT}^{-1}=\Delta/m(\Delta)$
(dots).
The ordered moment given by spin-DFT (GGA) is reproduced by 
$I_{\rm DFT}=0.82\,{\rm eV}$~\protect\cite{ole-lilia}.\label{fig:DFTmomentplot}}
\end{figure}

The local Hamiltonian $\hat{H}_{{\rm C},i}$ describes the Coulomb 
interaction between the $d$-electrons.
We use the Hubbard interaction~$U$ and an averaged Hund's-rule
interaction~$J$ to derive the Racah coefficients $A,B,C$ with the atomic ratio
$C/B=4$
which parameterize our multi-band interaction~\cite{schicklingbuenemann}.
In the following, until in Figs.~\ref{fig:magmomGT} 
and~\ref{fig:DFTmomentplot}, we give the results for a characteristic 
set of values $U=8\, {\rm eV}$ and $J=0.6\, {\rm eV}$ 
[$(U,J)=(8,0.6)\,{\rm eV}$]. This set, for which $J=0.075U$, 
gives an ordered moment of $m=0.74 \mu_{\rm B}$, intermediate 
between two recent experimental values. 

We fix the average number of 3$d$-electrons per Fe and of 
the 4$p$-electrons per As at their PM
DFT-values, $n_d=7.47$ and $n_p=4.53$, respectively~\cite{ole-lilia}.
The polar-covalent bonding between Fe and As
thus transfers 1.47 electrons from As to Fe
compared with the ionic picture $(d^6p^6)$ implicit in the often used 
five-band model~\cite{graser2009}.
It has been shown previously~\cite{schicklingbuenemann} that
the ionic picture does not lead to a satisfactory description
of the magnetically ordered phase of LaOFeAs within GT.
We do not fix the occupations of each of
the five 3$d$~orbitals but
there is little charge flow between them
as a function of~$(U,J)$ for
the parameter values considered.

The true ground state of $\hat{H}$ in~(\ref{1.1})
is approximated by the Gutzwiller variational wave function 
\begin{equation}
|\Psi_{\rm G}\rangle=\hat{P}_{\rm G}|\Psi_0\rangle
=\prod\nolimits_{i}\hat{P}_{i}|\Psi_0\rangle\;,
\label{1.3} 
\end{equation}
where $|\Psi_0\rangle$ is a product state of Bloch orbitals, 
and the local Gutzwiller correlator is defined as
\begin{equation}
\hat{P}_{i}=\sum\nolimits_{\Gamma}\lambda_{\Gamma}
|\Gamma \rangle_{i} {}_{i}\langle \Gamma |\;.
\label{1.4}
\end{equation}
Variational parameters $\lambda_{\Gamma}$ have been introduced
for each multiplet state $|\Gamma \rangle_{i}$ on the iron sites, 
i.e., the eigenstates of $\hat{H}_{{\rm C},i}$. 
Note that $|\Psi_0\rangle$
is also a variational object which we determine from the minimization 
of the variational energy functional resulting from 
the wave functions~(\ref{1.3}). 

As shown in Refs.~\cite{buenemann1998,buenemann2005}, 
expectation values can be evaluated without further approximations
in the limit of infinite spatial dimensions. 
The energy functional derived in this limit is then used 
as an approximation for finite-dimensional systems. 
In addition, GT provides
the Landau--Gutzwiller quasi-particle bandstructure
for comparison with ARPES data~\cite{buenemann2003b,hofmann2009-etal}.  

We begin with the hight-temperature 
PM phase and present the quasi-particle bandstructure 
in Fig.~\ref{fig:bandstructure}.
The electron-electron interaction
between the $d$-electrons is seen to reduce the overall bandwidth.
Filled bands with a large $3d$~character~\cite{ole-lilia} 
are shifted upwards in energy and empty, 3$d$-like bands downwards. 
Also, the three lowest bands are effected since of the 6~electrons
in these As~$p$-like bands, 1.82 are Fe~$d$.
\begin{figure}[htb]
\begin{tabular}{@{}ll@{}}
\includegraphics[height=3.6cm]{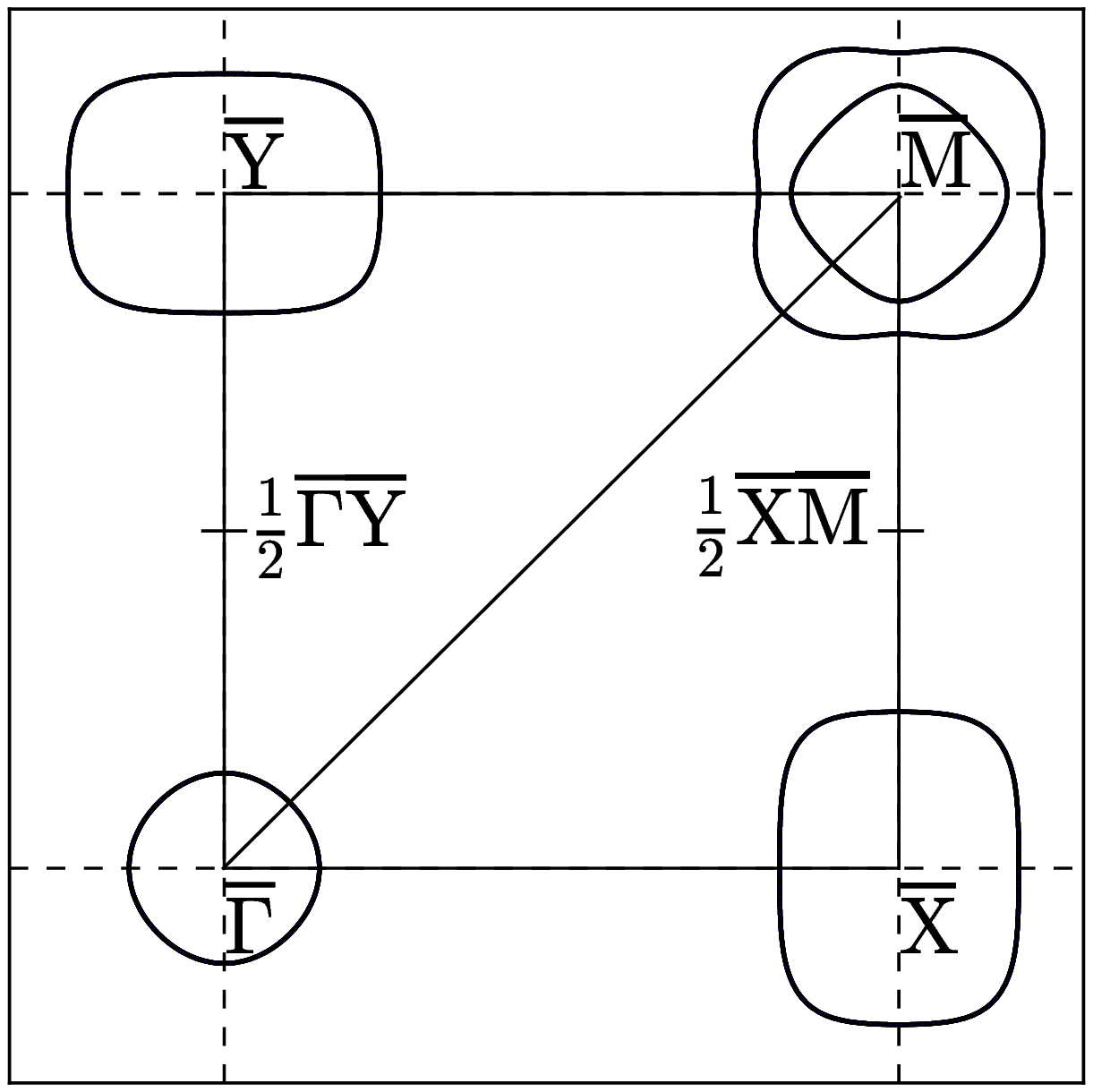} &
\includegraphics[height=3.6cm]{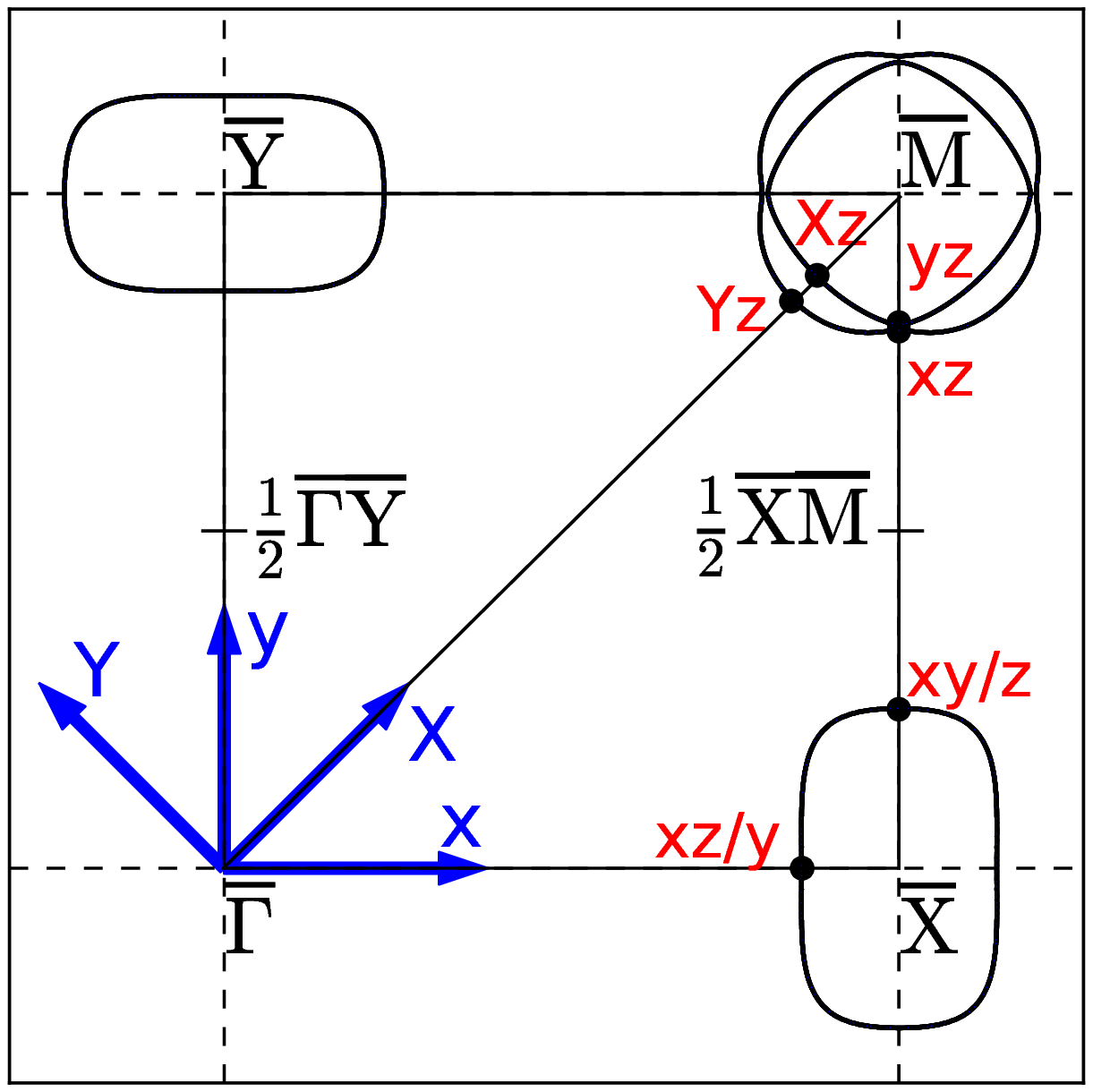}
\end{tabular}
\caption{(Color online) PM Fermi surface.
Left: DFT; Right: GT for $(U,J)=(8,0.6)\, {\rm eV}$.
The lower and upper case coordinate systems used 
to designate the orbitals are shown.\label{fig:fermisurface}}
\end{figure}

In Fig.~\ref{fig:fermisurface} we show the FS calculated with DFT and GT for 
$(U,J)=(8,0.6)\, {\rm eV}$ for the PM phase.
In GT the correlations shift the $xy$ band down by $130\, {\rm meV}$ 
and thereby empty the $\overline{\Gamma}$-centered 
hole pocket. The concomitant lowering of the Fermi level 
narrows the $\overline{\rm X}$  and $\overline{\rm Y}$-centered 
electron pockets and makes the inner $\overline{\rm M}$-centered hole pocket
expand.
The lowering of the empty $XY/Yz$-like band near $\overline{\rm M}$
steepens the lower, $Yz$-like conduction band~\cite{ole-lilia} 
and thereby shrinks the outer $\overline{\rm M}$-centered hole pocket 
along $\overline{\rm M}$-$\overline{\Gamma}$. 
The small size of these effects confirms
the conventional wisdom that DFT bands provide
a reasonable description of the FS. On the other hand,
the agreement of the DFT FS with experiment 
does not indicate that the electron-electron interactions are weak.

The correlation-induced mass renormalization on the FS is
defined as the renormalization of the Fermi velocity,
$m^*/m=v_{\rm F,DFT}/v_{\rm F}$. It
depends on the orbital character of the band states involved.
For pure $d$-bands, the renormalization is $q_d\approx 1.5$ for
$(U,J)=(8,0.6)\, {\rm eV}$ and grows by 3\% per eV~increase of $U$.
The mass enhancement decreases towards unity
with increasing $p$-character.
Also band-shifts influence the effective masses on the Fermi surface.
For the six points where the Fermi surface cuts
the high-symmetry lines 
in Fig.~\ref{fig:fermisurface},
$(m^*/m)_n=1.6,1.2,1.3,1.3,1.4,1.3$ for the orbitals
$n=xz/y,xy/z,xz,yz,Xz,Yz$.
Note that the $xz/y$-like DFT band in $\overline{\Gamma}$-$\overline{\rm X}$ 
direction
has a strong curvature in the vicinity of $\overline{\rm X}$
and, as the correlation-induced downwards shift of the $xy$ band 
drags the Fermi level along, 
it is placed lower in the $xz/y$ band. 
This increases $(m^*/m)_{xz/y}$ beyond the renormalization factor~$q_d$.

\begin{figure}[t]
\begin{tabular}{@{}ll@{}}
\includegraphics[width=4.1cm]{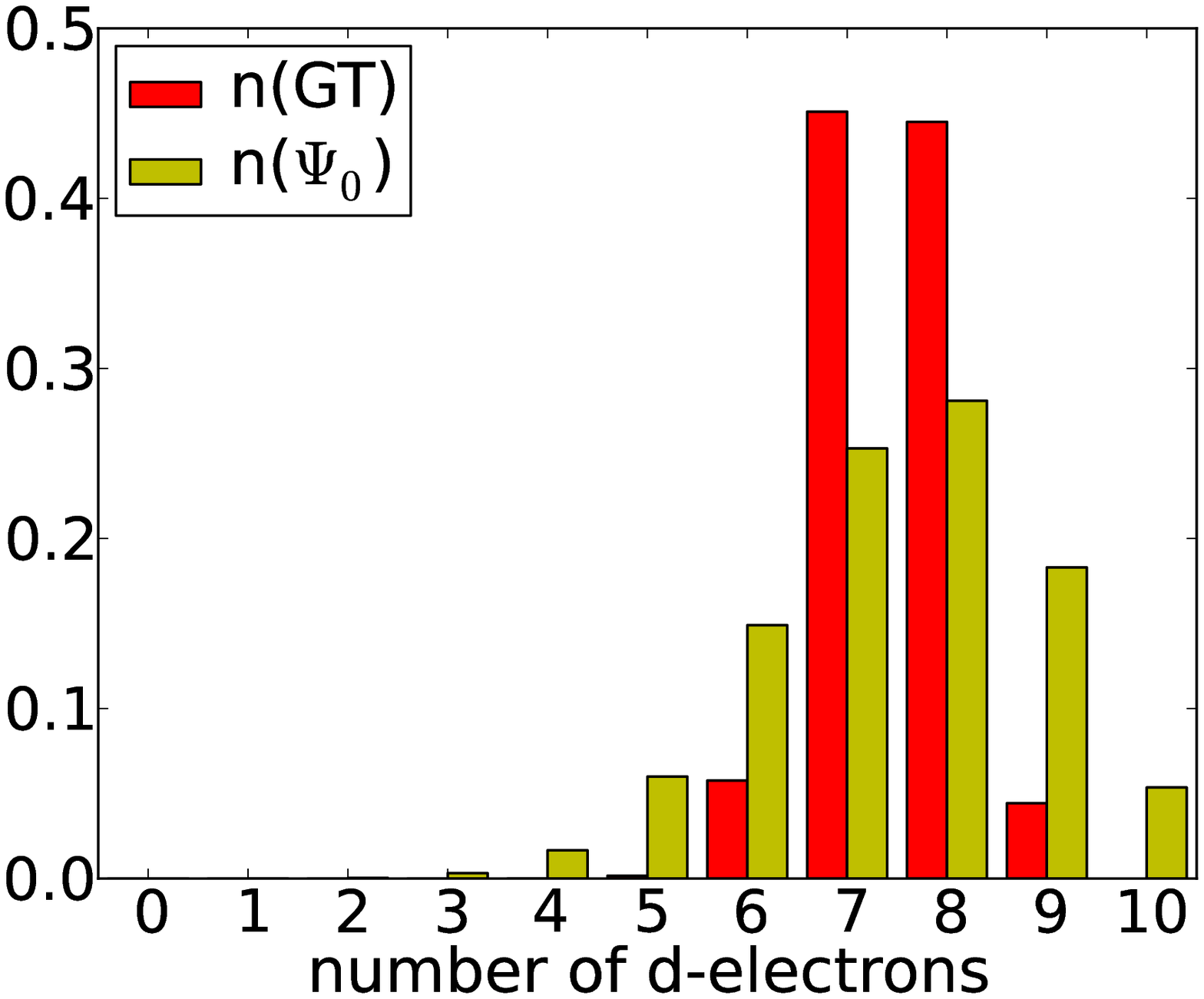} &
\includegraphics[width=4.1cm]{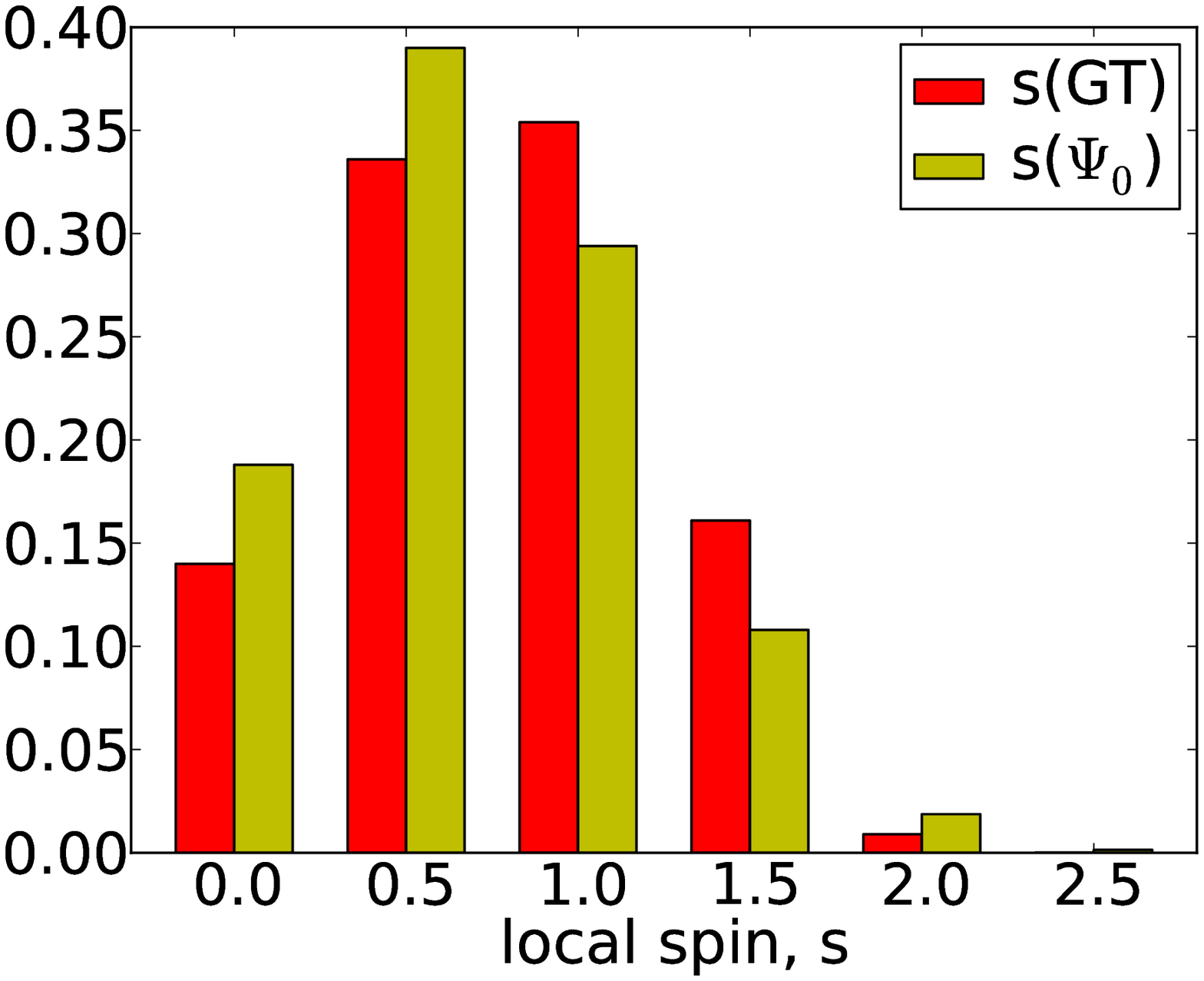}
\end{tabular}
\caption{(Color online) Local $d$-charge distribution $p(n)$ (left) and local
spin distribution $p(s)$ (right) for the PM GT 
with $(U,J)=(8,0.6)\,{\rm eV}$
and for the PM single-particle wave 
function $|\Psi_0\rangle$.\label{fig:localspinandcharge}}
\end{figure}

For $(U,J)=(8,0.6)\, {\rm eV}$,
the dominant 3$d$-con\-figur\-ations are 3$d^7$ and 3$d^8$
in the optimized PM Gutzwiller wave function $|\Psi_{\rm G}^{\rm opt}\rangle$.
Occupancies with $n_d=6$ or $n_d=9$ electrons are rare, and all other
charge states are essentially forbidden, see Fig.~\ref{fig:localspinandcharge}. 
The absence of large charge fluctuations
is typical for correlated electron systems.
The probability distribution function $p(s)$ for finding local spins with size
$0\leq s\leq 5/2$ in $|\Psi_{\rm G}^{\rm opt}\rangle$
is very similar to the distribution in the single-particle 
product state~$|\Psi_0\rangle$. In GT, the correlation enhancement of the
local spin moment is small. This shows that the system is far 
from the local-moment situation 
where we would solely find atoms with Hund's-rule spins $s=3/2$
for 3$d^7$ and $s=1$ for 3$d^8$. From 
the expectation values for the local spin, 
$\langle (\hat{\bf S}_i)^2\rangle =\sum_s p(s) s(s+1)$, we find
$\langle (\hat{\bf S}_i)^2\rangle =1.62$ and 
$\langle (\hat{\bf S}_i)^2\rangle_0 =1.41$ 
for $|\Psi_{\rm G}^{\rm opt}\rangle$
and $|\Psi_0\rangle$, respectively, considerably smaller than
$\langle (\hat{\bf S}_i)^2\rangle_{s=3/2} =15/4$ for $s=3/2$ and 
$\langle (\hat{\bf S}_i)^2\rangle_{s=1} =2$ for $s=1$.

Now, we turn to the SDW.
The local charge and spin distributions shown 
in Fig.~\ref{fig:localspinandcharge} change only slightly 
from the PM to the SDW phase, as also seen in DMFT~\cite{kotliarNPL}.
The exchange splitting, $\Delta$, is within 10\% the same for all five
$d$-orbitals and amounts to $\Delta=0.37\, {\rm eV}$ 
for $(U,J)=(8,0.6)\, {\rm eV}$.
The $d_{yz}$, $d_{xy}$, $d_{3z^2-r^2}$, and $d_{x^2-y^2}$ orbitals contribute 
almost equally to the magnetization, which is $m=0.74 \mu_{\rm B}$ 
for $(U,J)=(8,0.6)\, {\rm eV}$,
whereas the $d_{xz}$ orbital, 
whose band hardly disperses in the anti-ferromagnetic direction, 
is twice as polarized as the others.
Its exchange splitting is 10\% larger.

On the right-hand side of Fig.~\ref{fig:bandstructure},
we compare the quasi-particle bandstructures of
the PM and SDW phases for $(U,J)=(8,0.6)\, {\rm eV}$.
Where the PM $E(\kvec)$-bands cross
the $E(\kvec+\qvec)$ bands 
($\qvec=[\overline{\Gamma}\,\overline{\rm Y}]=[\overline{\rm X}\,
\overline{\rm M}]$)
they split approximately by $\Delta$ times the overlap of their $d$-orbital
characters~\cite{ole-lilia}. This leads 
to a noticeable rearrangement of the bands.
In GT also band-shifts are permitted.
Nevertheless, the GT quasi-particle bands and FS
in the SDW phase are similar to the results of Stoner theory
using the PM DFT bands from $\hat{H}_0$ in~(\ref{1.1})
and the exchange-interaction constant~$I_{\rm Stoner}$
adjusted to give the same moment~\cite{ole-lilia}.

The dependence of the ordered
magnetic moment on the atomic parameters~$(U,J)$
in GT is shown in Fig.~\ref{fig:magmomGT}.
We find a remarkably broad region in the $(U,J)$ parameter space with a small
ordered moment $m\lesssim 0.8\mu_{\rm B}$,
in contrast to a description based on six~Fe electrons
in five orbitals~\cite{graser2009,schicklingbuenemann}.
As for the five-band model~\cite{schicklingbuenemann}, 
the PM metal is the ground state for not too large values
of the Hund's-rule coupling~$J$, 
in contrast to Hartree--Fock theory where symmetry breaking
occurs for small~$U$.

The atomic Hamiltonian
is frequently approximated by density-density interactions only,
e.g. in DMFT~\cite{aichhorn-prb2009-etal,hansmann-etal}
in order to keep the numerical problems under control.
One of the magnetization curves in Fig.~\ref{fig:magmomGT} compares the 
GT result of this approximation with the GT for the full Hamiltonian.
We see that the former approximation largely overestimates the magnetization.

The inset of Fig.~\ref{fig:magmomGT} shows that
the onset of the SDW requires 
a finite~$J\geq J_{\rm c}(U)\approx I_{\rm c}-0.017U$ 
with $I_{\rm c}=0.52\, {\rm eV}$ ($U\geq 3\, {\rm eV}$).
Moreover, in the SDW region, the moment is to a very good approximation 
a function of merely the linear combination $J+0.017U$. 
Thus, we then can identify 
$I(U,J)=J+0.017U$ as the effective low-energy
scale for the magnetic excitations of our eight-band model in GT.
That not only the onset but
the entire GT magnetization collapses onto one curve
when plotted as a function of~$I$ is seen in Fig.~\ref{fig:DFTmomentplot}. 
This shows that the Hund's-rule coupling and not
the Hubbard-repulsion is the controlling parameter
for the magnetism in this system.
This may explain why DMFT studies of LaOFeAs obtain a small ordered moment
using quite different 
$U$-values~\cite{aichhorn-prb2009-etal,hansmann-etal,kotliarNPL}.

Using the PM DFT Hamiltonian~$\hat{H}_0$, eq.~(\ref{1.1}),
we calculate the magnetization, $m(\Delta)$, as a function of  
an external staggered field~$\Delta$.
The resulting $\Delta/m(\Delta)\equiv \chi_{\rm DFT}^{-1}$
is a pure bandstructure function, which in Stoner theory 
provides the selfconsistent magnetization as the solution of
the equation $\chi_{\rm DFT}^{-1}(m)=I_{\rm Stoner}$~\cite{ole-lilia}.
This function is plotted in Fig.~\ref{fig:DFTmomentplot}.
As seen from the size of the magnetization jump,
the transition from the PM to the SDW state in GT is driven by
the same band mechanism. Details differ
due to different underlying PM bandstructures.

The existence of a sharp `nose' at small moments is caused by the nesting 
of the hole and electron sheets with common $xz$ character, while 
the strong increase of the moment, once $m\gtrsim 0.75 \mu_{\rm B}$ in GT,
occurs when the $xy$ exchange splitting is so large that the 
spin-$\uparrow$ minority band empties at $\overline{\Gamma}=\overline{\rm Y}$,
and thereby 
starts to contribute to the magnetization.   
At this point also the $zz$ magnetization picks up in GT. 
This is mainly due to $zz$ hybridization of the empty $XY$ minority band near 
$\overline{\rm Y}$. Nevertheless, over the entire parameter range, 
$m_{xz}>m_{xy}>m_{zz}>m_{XY}>m_{yz}$ in GT. For DFT 
the moment increases faster after the nose and 
$m_{xz}=m_{xy}>m_{zz}>m_{XY}>m_{yz}$. 
The first difference is due to the steeper bands 
and the latter to the slightly different band positions. Roughly speaking,
the GT of correlated quasi-particles
supports the Stoner picture of band magnetism 
for the small-moment phase in LaOFeAs, provided that 
the DFT exchange-correlation
kernel is substituted by a weak, low-energy Stoner parameter.

The correlated band picture can be further scrutinized
from the analysis of the GT
susceptibility $\chi_{\rm GT}=\Delta_{\rm GT}/m(\Delta_{\rm GT})$ (not shown).
Following Landau Fermi-liquid theory, we write it as
$\chi_{\rm GT}/\chi_{\rm DFT}= q_d/(1+F_0^{\rm a})$ where $q_d$ is
the average mass enhancement factor 
and $F_0^{\rm a}$ is a Landau parameter. 
The enhancement of the susceptibility is mostly due 
to~$q_d(U,J)$, i.e., $\chi_{\rm GT}/q_d$ depends
on the interaction through~$I(m)$ only,
thus explaining why the three GT curves in Fig.~\ref{fig:DFTmomentplot}
can be rescaled to a universal curve
even though their mass enhancements are very different.
For all $m$
we find that $\chi_{\rm GT}/(q_d\chi_{\rm DFT})=1/(1+F_0^{\rm a}(m))\approx 0.8$.
A small positive value $F_0^{\rm a}(m)\approx 0.2$
reflects the tendency to anti-ferromagnetic spin alignment.
The applicability of Fermi-liquid theory supports the
DFT picture of band magnetism in LaOFeAs.

In this work, we have employed Gutzwiller theory
to calculate the quasi-particle bands for a two-dimensional
eight-band Hubbard model for the valence
electrons in the iron-arsenic planes of LaOFeAs. 
In a large region of the $(U,J)$ parameter space,
we find a spin-density wave ground state of quasi-particles
with a small magnetic moment $m\lesssim 0.8\mu_{\rm B}$, 
as observed experimentally.
When charge fluctuations are suppressed by the Hubbard-$U$, 
the Hund'-rule coupling~$J$ 
is the effective low-energy scale which drives the transition
to metallic band magnetism.

In view of the qualitatively correct DFT description 
of the paramagnetic and also the spin-density wave phases (allowing
for a renormalization of
$I_{\rm DFT}\sim 0.8\, {\rm eV}$ to $I_{\rm Stoner}\sim 0.6\, {\rm eV}$),
the DFT bandstructure provides
a reasonable starting point for investigations
of the iron-based superconductors.
The Gutzwiller theory offers a microscopic Fermi-liquid description
for the correlation corrections due to the electron-electron
interaction.

\begin{acknowledgments}
This work was supported in part by the 
Deutsche Forschungsgemeinschaft (SPP 1458 and Bo-3536/1).
\end{acknowledgments}

\bibliographystyle{unsrt}
\bibliography{gebhard}

\begin{thebibliography}{10}

\bibitem{kamihara2008}
Y.~Kamihara, T.~Watanabe, M.~Hirano, and H.~Hosono.
\newblock {\em J. Am. Chem. Soc.}, 130:3296, 2008.

\bibitem{magmom1-etal}
C.~de~la Cruz~{\sl et al.}
\newblock {\em Nature~(London)}, 453:899, 2008.

\bibitem{magmom2-etal}
N.~Qureshi {\sl et al.}
\newblock {\em Phys.~Rev.~B}, 82:184521, 2010.

\bibitem{magmom3-etal}
H.~F.~Li {\sl et al.}
\newblock {\em Phys.~Rev.~B}, 82:064409, 2010.

\bibitem{FESC:review}
J.~Paglione and R.~L. Greene.
\newblock {\em Nature physics}, 6:645, 2010.

\bibitem{FESC:review-2}
D.~C. Johnston.
\newblock {\em Adv. in physics}, 59:803, 2010.

\bibitem{magmomdft1-lilia}
D.~J. Singh and M.-H. Du.
\newblock {\em Phys.~Rev.~Lett.}, 100:237003, 2008.

\bibitem{ole-lilia}
O.~K. Andersen and L.~Boeri.
\newblock {\em Ann. Physik (Berlin)}, 523:8--50, 2011.

\bibitem{magmomdft2-lilia-etal}
I.~I.~Mazin {\sl et al.}
\newblock {\em Phys.~Rev.~Lett.}, 101:057003, 2008.

\bibitem{aichhorn-prb2009-etal}
M.~Aichhorn {\sl et al.}
\newblock {\em Phys.~Rev.~B}, 80:085101, 2009.

\bibitem{kotliarNPL}
Z.~P. Yin, K.~Haule, and G.~Kotliar.
\newblock {\em Nature physics}, 7:294, 2011.

\bibitem{hansmann-etal}
P.~Hansmann {\sl et al.}
\newblock {\em Phys. Rev. Lett.}, 104:197002, 2010.

\bibitem{schicklingbuenemann}
T.~Schickling, F.~Gebhard, and J.~B{\"u}nemann.
\newblock {\em Phys. Rev. Lett.}, 106:146402, 2011.

\bibitem{graser2009}
S.~Graser, T.~Maier, P.~Hirschfeld, and D.~Scalapino.
\newblock {\em New Journal of Physics}, 11:025016, 2009.

\bibitem{buenemann1998}
J.~B{\"u}nemann, W.~Weber, and F.~Gebhard.
\newblock {\em Phys.~Rev.~B}, 57:6896, 1998.

\bibitem{buenemann2005}
J.~B{\"u}nemann, F.~Gebhard, and W.~Weber.
\newblock In A.~Narli-kar, editor, {\em Frontiers in Magnetic Materials}, pages
  117--151. Springer, Berlin, 2005.

\bibitem{buenemann2003b}
J.~B{\"u}nemann, F.~Gebhard, and R.~Thul.
\newblock {\em Phys.~Rev.~B}, 67:75103, 2003.

\bibitem{hofmann2009-etal}
A.~Hofmann {\sl et al.}
\newblock {\em Phys.~Rev.~Lett.}, 102:187204, 2009.

\end{thebibliography}

\end{document}